# Anisotropic Thermal Characterization of Suspended and Spin-Coated Polyimide Films Using a Square-Pulsed Source Method

Bingjiang Zhang[1], Dihui Wang[2], Tao Chen[1], Heng Ban[2,*], Puqing Jiang[1,*]

[1]*School of Energy and Power Engineering, Huazhong University of Science and Technology, Wuhan, Hubei 430074, China*

[2]*Swanson school of engineering, University of Pittsburgh, Pittsburgh, US*

**ABSTRACT:** Polyimide (PI) thin films are widely used in advanced technologies, yet accurate characterization of their thermal properties remains challenging, as evidenced by significant inconsistencies in reported data and an incomplete understanding of heat transfer mechanisms. In this study, we employ an optical Square-Pulsed Source (SPS) technique to simultaneously measure the in-plane and cross-plane thermal conductivities, as well as the volumetric heat capacity, of PI thin films. SPS is a pump-probe method that utilizes a square-wave-modulated pump laser to induce periodic heating and a probe laser to detect the thermoreflectance response. Thermal properties are extracted by analyzing amplitude signals across multiple modulation frequencies and laser spot sizes. Measurements were conducted on both suspended commercial PI films and spin-coated PI films on fused silica substrates. The results show that spin-coated films exhibit higher cross-plane thermal conductivity and lower anisotropy compared to suspended films, which we attribute to differences in molecular orientation and substrate interactions. These findings provide new physical insights into anisotropic heat transport in polymer thin films and demonstrate the SPS technique as a robust tool for probing microscale thermal phenomena in soft materials.

*Corresponding Author: hengban@pitt.edu (H.B.); jpq2021@hust.edu.cn (P.J.)

# 1. Introduction

Polyimide (PI) thin films are widely used in advanced applications such as microelectronics, aerospace engineering, and energy systems due to their exceptional thermal stability, chemical resistance, and mechanical strength [1–5]. These attributes make PI films ideal for high-temperature circuits, flexible electronics, and thermal insulation in spacecraft. However, as devices' miniaturization and power densities continue to increase, effective thermal management has become critical, underscoring the need for accurate thermal property characterization to ensure reliable performance.

Thermal transport in PI thin films is inherently anisotropic, largely influenced by molecular alignment and processing conditions [6–9]. Typically, the in-plane thermal conductivity ($k_r$) is significantly higher than the cross-plane conductivity ($k_z$), which has important implications for heat dissipation in electronic devices. Although advanced techniques such as Structured Illumination with Thermal Imaging (SI-TI) [10] and Spatially-Resolved Lock-in micro Thermography (SR-LIT) [11,12] have been developed to measure anisotropic thermal conductivity, they are optimized for bulk materials and face challenges when applied to micron-scale thin films. Moreover, these approaches often require multiple experimental configurations to extract different thermal parameters, increasing measurement complexity and introducing potential inconsistencies.

To overcome these challenges, several methods tailored for thin film characterization have been developed. Kurabayashi et al. [9] introduced three techniques to independently measure $k_r$ and $k_z$ of spin-coated PI films (0.5-2.5 μm thick). Two methods utilized micromachined structures with steady-state Joule heating and resistance thermometry, while a third, IC-compatible approach enabled simultaneous extraction of both components. Their results revealed pronounced anisotropies, with typical values of $k_r \approx 1.2$ W m$^{-1}$ K$^{-1}$ and $k_z \approx 0.25$ W m$^{-1}$ K$^{-1}$ for films 2-2.5 μm thick. Notably, $k_z$ was determined without requiring knowledge of the volumetric heat capacity ($C$), while the evaluation of $k_r$ relied on lateral thermal diffusivity measurements using a reported $C$ value of 1.55 MJ m$^{-3}$ K$^{-1}$ from DuPont data.

Takahashi et al. [13] employed the ac-calorimetric distance-variation method and temperature wave analysis to measure the anisotropic thermal diffusivities of two PI types (UPILEX-S and UPILEX-R) with thicknesses of 7.5-20.5 μm. They observed that the in-plane thermal diffusivity ($\alpha_r$) of UPILEX-R (standard polyimide) decreased with increasing thickness from $5.54 \times 10^{-7}$ m² s⁻¹ (7.5 μm) to $3.47 \times 10^{-7}$ m² s⁻¹ (20.5 μm), while the cross-plane thermal diffusivity ($\alpha_z$) remained relatively constant ($1.21 - 1.27 \times 10^{-7}$ m² s⁻¹). For UPILEX-S (advanced polyimide), $\alpha_r$ was significantly higher (7.32–9.02× $10^{-7}$ m² s⁻¹), attributed to enhanced molecular chain alignment and tensile strength, whereas $\alpha_z$ (1.01–1.30× $10^{-7}$ m² s⁻¹) remained insensitive to film properties.

More recently, Zhang et al. [14] developed a laser spot periodic heating method with sub-region phase fitting to determine anisotropic thermal diffusivities in thin films, reporting a notably high anisotropy ratio of 12 ($k_r = 1.2$ W m⁻¹ K⁻¹, $k_z = 0.1$ W m⁻¹ K⁻¹) for an 8 μm PI film. Chowdhury et al. [15] employed displacement thermo-optic phase spectroscopy (D-TOPS) and time-domain thermoreflectance (TDTR) to characterize Kapton PI films, reporting $k_r$ values of $0.62 - 0.64$ W m⁻¹ K⁻¹ and $k_z$ values of $0.17 - 0.21$ W m⁻¹ K⁻¹ for unfilled PI films (EN/HN), with anisotropy ratios varying from 2 to 4. Kim et al. [16] measured the cross-plane thermal conductivity of a colorless polyimide (CPI) using the micro time-domain thermoreflectance (μTDTR) technique, reporting a value of $k_z = 0.253$ W·m⁻¹·K⁻¹. However, many of these methods rely on assumed or separately measured heat capacities, which can introduce significant uncertainties.

Despite these advances, the literature reveals wide variability in reported thermal conductivities and anisotropy ratios, stemming from differences in measurement techniques, film thicknesses, molecular structures, and processing histories. Furthermore, few studies have directly measured the volumetric heat capacity of PI films. Diaham et al.[17] determined the specific heat capacity ($c_p = 1.09$ kJ kg⁻¹ K⁻¹) via differential scanning calorimetry (DSC) and estimated $C = 1.613$ MJ m⁻³ K⁻¹ using density measurements. However, such values can vary depending on film formulation and processing conditions, emphasizing the need for more comprehensive characterization methods.

In summary, while steady-state methods offer direct thermal conductivity measurements, they are less effective for thin films due to low thermal conductivity and small thicknesses. Transient techniques such as AC calorimetry and TDTR offer higher precision but typically require knowledge of the heat capacity, potentially limiting their accuracy. Therefore, there remains a critical need for an integrated, high-resolution method capable of simultaneously determining anisotropic thermal conductivities and volumetric heat capacity in thin polymer films.

To address this need, we introduce a non-contact Square-Pulsed Source (SPS) method for simultaneous measurement of in-plane and cross-plane thermal conductivities and volumetric heat capacity of thin films within a single experimental setup. The SPS method employs square-wave heating over a broad modulation frequency range (1 Hz to 10 MHz), combined with high-precision amplitude signal detection, ensuring robust and reliable measurements. In comparison, other thermoreflectance techniques, such as TDTR, are limited by a lowest measurable frequency of 0.1 MHz, which restricts their ability to accurately measure low in-plane thermal conductivity [18]. FDTR, on the other hand, employs sinusoidal modulation of the pump laser and relies on frequency-dependent phase signals for parameter fitting. However, as the frequency decreases, these phase signals approach zero, making low-frequency signals ineffective for accurate parameter extraction [19]. In contrast, the SPS method is particularly suited for characterizing thin films with low thermal conductivity, offering greater accuracy, enhanced sensitivity at low frequencies, and improved experimental flexibility.

In this study, we apply the SPS method to characterize three suspended PI thin films (Kapton HN-100 and EN-100, and a 15 μm Kaneka PI film) and three spin-coated PI films on fused silica substrates (4.45, 5.0, and 10 μm thick). Our results show that $k_r$ is similar for both film types, while $k_z$ is higher in the spin-coated films, indicating structural differences when compared to the commercial products. The measured volumetric heat capacities are consistent across all samples, demonstrating the accuracy and robustness of the SPS technique.

**2. Methodologies**

*2.1. Fundamental principle of the SPS method*

As illustrated in Fig. 1, the SPS method utilizes a TEM$_{00}$ pump laser modulated with a 50% duty-cycle square wave to induce periodic heating on the sample surface. A probe laser, concentrically aligned with the pump beam, monitors the resulting temperature variations via the thermoreflectance effect. The reflected probe beam is collected by a photodetector, and the signals are processed using a periodic waveform analyzer integrated within a Zurich Instruments UHF lock-in amplifier, enabling high signal-to-noise ratio extraction of voltage waveforms over the entire square-wave heating cycle. The measured waveforms are subsequently normalized in amplitude and time and fitted with a thermal model to extract the thermal properties of the sample [20,21].

As a thermoreflectance method, SPS typically requires the deposition of a ~100 nm metal transducer layer to facilitate thermoreflectance detection. At the probe wavelength of 785 nm, materials such as aluminum (Al), titanium nitride (TiN), and hafnium nitride (HfN) are commonly used due to their high thermoreflectance coefficients. In addition to thermoreflectance performance, the selection of transducer materials is influenced by their thermal conductivity and optical absorptance at the probe wavelength.

Transducers with a low thermal conductivity are particularly advantageous for resolving the in-plane thermal diffusivity of materials with thermal conductivities below 1 W m$^{-1}$ K$^{-1}$ [22]. For this reason, magnetron-sputtered Al (~50 W m$^{-1}$ K$^{-1}$) is preferred over thermally evaporated Al (~200 W m$^{-1}$ K$^{-1}$) for measuring the in-plane thermal diffusivity of PI films. Additionally, low optical absorptance at the probe wavelength helps minimize probe-induced steady-state heating, which is especially important when measuring thermally insulating materials like PI. Given these considerations, magnetron-sputtered Al is identified as the most suitable transducer material for SPS measurements on low-conductivity thin films. Although the thermoreflectance coefficient of Al films may vary depending on the deposition method, normalizing the experimental signals effectively eliminates the influence of these variations.

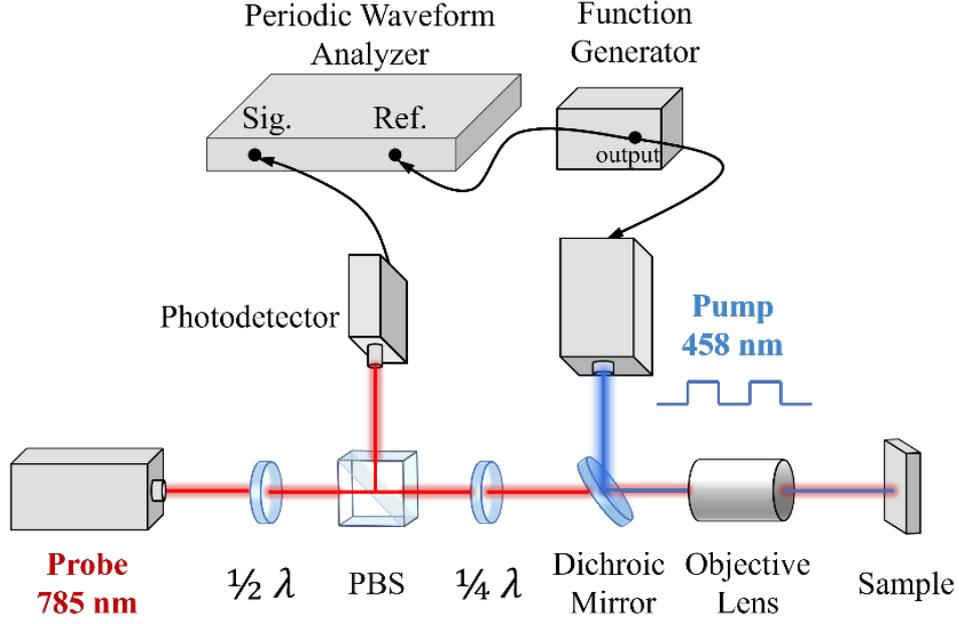

**Fig. 1.** Schematic diagram of the SPS experimental setup.

*2.2. Principle of simultaneously measuring anisotropic thermal conductivities and heat capacity of thin films*

When measuring a thin film sample in thermoreflectance experiments, parameters governing the heat diffusion process include $k_r$, $k_z$, $C$, and $h$ of the film, where $h$ is the film thickness. Chen and Jiang [23] demonstrated that these parameters are fundamentally interrelated through three combined parameters: $\frac{\sqrt{k_z C}}{h_{Al} C_{Al}}$, $\frac{\sqrt{k_z C}}{hC}$, and $\frac{k_r}{Cr_0^2}$, where $h_{Al}$ and $C_{Al}$ represent the thickness and volumetric heat capacity of the metal transducer layer, respectively, and $r_0$ is the root-mean-square average of the $1/e^2$ radii of the pump and probe laser spots. Given the known values of $h_{Al}$, $C_{Al}$, $h$, and $r_0$, the thermal properties $k_r$, $k_z$, and $C$ of the thin film sample can be simultaneously determined from these composite parameters.

It is important to note that simultaneous determination of anisotropic thermal conductivities and heat capacity is only possible when the sample is a thin film of finite thickness. In such cases, thermoreflectance experiments effectively determine the cross-plane thermal effusivity ($\sqrt{k_z C}$), in-plane thermal diffusivity ($k_r/C$), and areal heat capacitance ($hC$) of the film. Knowing the film thickness then allows decoupling and simultaneous extraction of $k_r$, $k_z$, and $C$. In contrast, for semi-infinite samples, only $\sqrt{k_z C}$ and $k_r/C$ can be determined, making it impossible to independently resolve $k_r$, $k_z$, and $C$.

The sensitivity of the normalized amplitude signal $A_{\text{norm}}$ to a parameter $\xi$ is quantified by the sensitivity coefficient:

$$S_\xi = \frac{\partial \ln A_{\text{norm}}}{\partial \ln \xi} = \frac{\xi}{A_{\text{norm}}} \frac{\partial A_{\text{norm}}}{\partial \xi} \tag{1}$$

This coefficient implies that a 1% change in $\xi$ leads to an $S_\xi$% change in $A_{\text{norm}}$.

The sensitivity coefficients of the combined thermal parameters can be related to those of the individual parameters as follows [23]:

$$S_{\frac{\sqrt{k_z C}}{hC}} = 2S_{k_z} + 2S_{k_{z,\text{Al}}} + S_{h_{\text{Al}}} + S_G \tag{2}$$

$$S_{\frac{\sqrt{k_z C}}{h_{\text{Al}} C_{\text{Al}}}} = -\left(2S_{k_{z,\text{Al}}} + S_{h_{\text{Al}}} + S_G\right) \tag{3}$$

$$S_{\frac{k_r}{C r_0^2}} = S_{k_r} \tag{4}$$

where $k_{z,\text{Al}}$ is the cross-plane thermal conductivity of the metal transducer layer, and $G$ is the interfacial thermal conductance between the transducer and the sample.

These expressions establish a systematic framework for simultaneously extracting $k_r$, $k_z$, and $C$ of a film sample from the thermal response signals measured using the SPS method.

**3. Results and discussion**

This study examines the thermal properties of six PI films, including three suspended films: Kapton HN-100, EN-100 (both 25 μm thick), and Kaneka (15 μm thick), as well as three spin-coated films on fused silica substrates with thicknesses of 4.45 μm, 5.0 μm, and 10 μm. To illustrate the measurement procedures and data analysis approach, we present detailed case studies of two representative samples: the 25 μm-thick suspended Kapton HN-100 film and the 5.0 μm-thick spin-coated PI film. A comparative summary of the measured thermal properties for all samples is provided at the end of this section.

*3.1. Thermal characterization of suspended PI film structures*

Figure 2(a) shows the sample structure used to measure suspended PI films. A ~100 nm-thick Al transducer layer was deposited on the PI surface, which was modeled as being supported by a semi-infinite air substrate. The corresponding thermal model includes eleven parameters: $k_{Al}$, $C_{Al}$, and $h_{Al}$ of the Al layer; $k_r$, $k_z$, $C$, and $h$ of the PI film; the interfacial thermal conductance $G$ between Al and PI; the thermal conductivity and volumetric heat capacity of air ($k_{air}$ and $C_{air}$); and the laser spot radius ($r_0$). Note that while the Al film is isotropic with $k_{Al} = k_{r,Al} = k_{z,Al}$, this study considers only $k_{r,Al}$, as the measurement signals are not sensitive to $k_{z,Al}$.

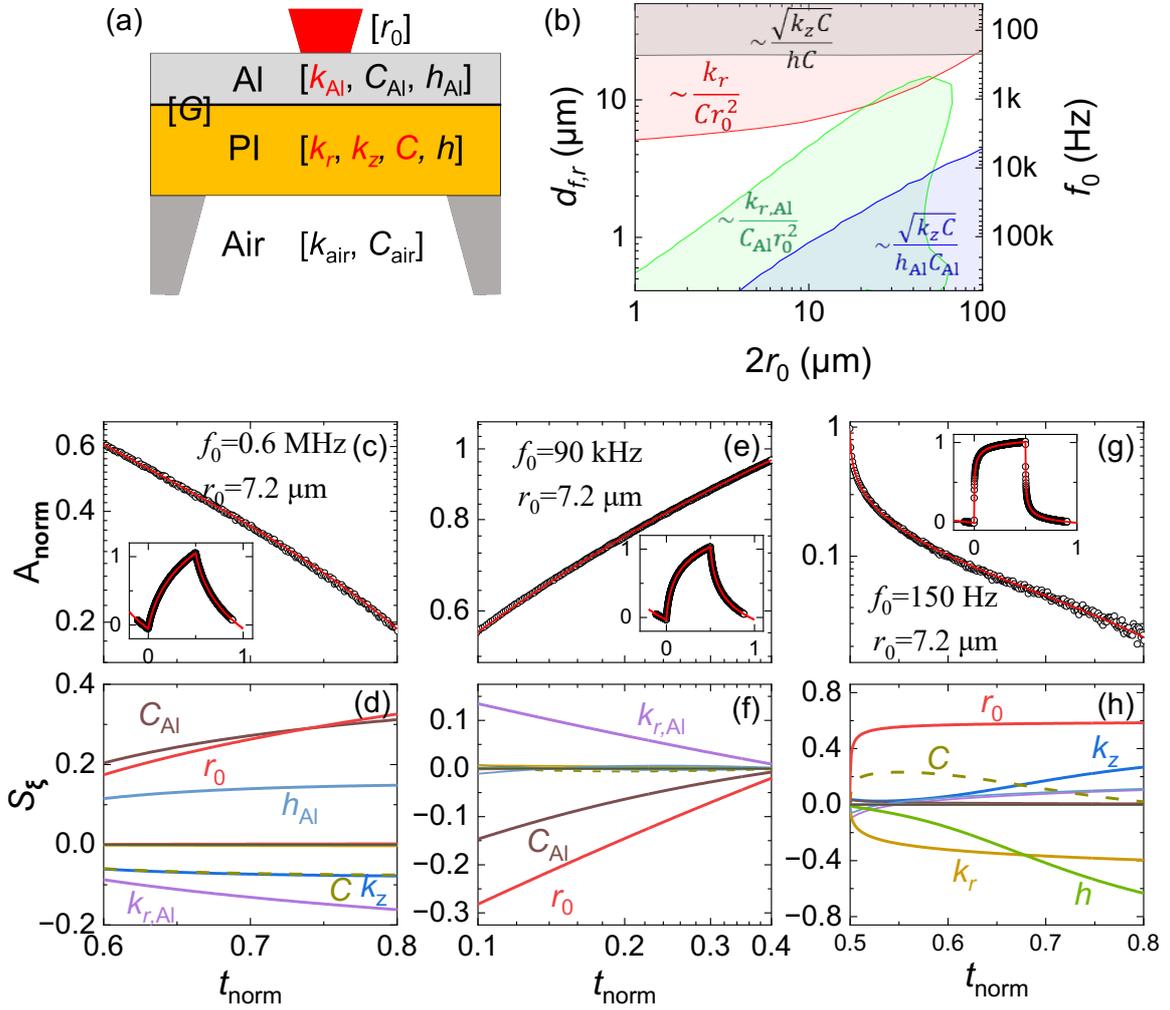

**Fig. 2.** (a) Schematic of the suspended PI sample structure used for thermal property measurements. A ~100 nm-thick Al transducer layer is deposited on the PI film, which is suspended over an air gap. (b) Sensitivity coefficients of thermoreflectance signal to key parameter combinations, plotted as functions of laser spot diameter and in-plane thermal diffusion length, guiding frequency selection for decoupled extraction. Regions with sensitivity magnitudes exceeding 0.2 are highlighted. (c-h) Measured SPS signals of the Kapton HN-100 PI film as functions of normalized time ($t_{norm}$) at modulation frequencies of (c, d) 0.6 MHz, (e, f) 90 kHz, and (g,h) 150 Hz, along with corresponding sensitivity analyses for key parameters. The insets in panels (c), (e), and (g) show the fitting of the signals over the full periodic heating cycle.

Among the eleven parameters, $k_{air}$, $C_{air}$, and $G$ have negligible influence on the measurement signals and are therefore fixed at $k_{air} = 0.026$ W m$^{-1}$ K$^{-1}$, $C_{air} = 1$ kJ m$^{-3}$ K$^{-1}$, and $G = 100$ MW m$^{-2}$ K$^{-1}$ as input parameters. A large uncertainty of $\pm 50\%$ was assumed for each of these parameters in the uncertainty analysis, which demonstrated that their uncertainties have a negligible impact on the extracted properties of the PI films.

Four parameters must be predetermined with high accuracy to serve as reliable inputs: $C_{Al}$, $h_{Al}$, $h$, and $r_0$. The volumetric heat capacity $C_{Al}$ is adopted from the literature [24] as $2.44 \pm 0.07$ MJ m$^{-3}$ K$^{-1}$. The PI film thicknesses were measured using a step profiler, yielding $15.2 \pm 0.2$ μm (Kaneka), $24.4 \pm 0.2$ μm (Kapton HN-100), and $25.0 \pm 0.2$ μm (Kapton EN-100).

A differential calibration approach was used to determine $h_{Al}$ and $r_0$. During Al deposition, a standard fused silica sample was placed adjacent to each test sample to ensure identical $h_{Al}$. Standard SPS measurements on fused silica at 1 MHz determines $\sqrt{k_{z,silica} C_{silica}}/(h_{Al} C_{Al})$, as shown in Fig. 3(a, b). Using known values of $k_{silica} = 1.38$ W m$^{-1}$ K$^{-1}$ [25] and $C_{silica} = 1.65$ MJ m$^{-3}$ K$^{-1}$ [26], $h_{Al}$ was determined to be 105 nm, with an estimated uncertainty of 3%. Similarly, SPS measurements at 2 kHz enabled the determination of $k_{r,silica}/(C_{silica} r_0^2)$, as shown in Fig. 3(c, d), from which $r_0$ is determined as 7.2 μm, with an uncertainty of 2%. A high-resolution microscope was used during experiments to precisely position each sample at the objective lens's focal plane, ensuring a consistent laser spot size across all measurements.

The remaining four parameters, namely $k_{Al}$, $k_r$, $k_z$, and $C$, are simultaneously determined through multi-frequency SPS measurements. These parameters are embedded within four composite quantities: $k_{r,Al}/(C_{Al} r_0^2)$, $k_r/(C r_0^2)$, $\sqrt{k_z C}/(h_{Al} C_{Al})$, and $\sqrt{k_z C}/(hC)$. By fitting these composite parameters and combining them with the predetermined input values, the individual thermal properties of the Al layer and PI films can be simultaneously extracted.

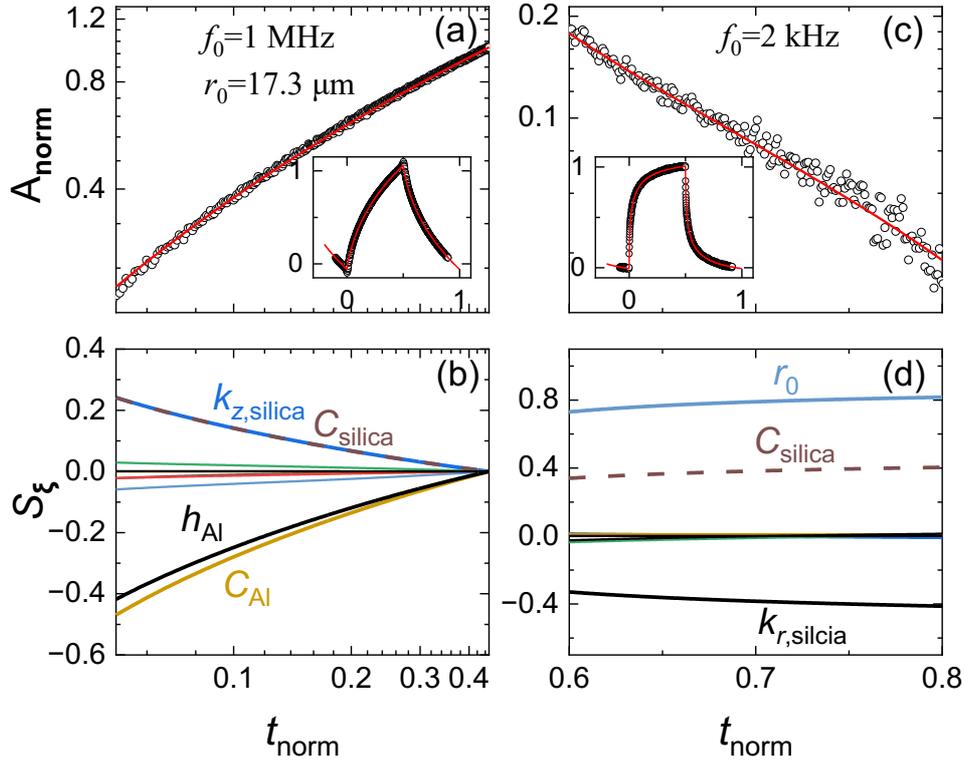

**Fig. 3.** SPS signals and sensitivity analysis of the standard fused silica sample for determining the Al film thickness ($h_{Al}$) and laser spot radius ($r_0$).
(a) SPS signals measured at a modulation frequency of 1 MHz with a spot size of 17.3 μm.
(b) Sensitivity coefficients of the signals in (a) to all parameters of the Al/silica system.
(c) SPS signals measured at a modulation frequency of 2 kHz with the same spot size.
(d) Sensitivity coefficients of the signals in (c) to all parameters of the Al/silica system.
The insets in panels (a) and (c) show the fitting of the signals over the full periodic heating cycle.

The sensitivity of each composite parameter varies with the heat conduction regime, which depends on the in-plane thermal diffusion length $d_{f,r} = \sqrt{k_r/(C\pi f)}$ and laser spot diameter $2r_0$ [21]. To guide experimental optimization, sensitivity coefficients as functions of $d_{f,r}$ and $2r_0$ are plotted in Fig. 2(b), with regions of high sensitivity (magnitude >0.2) highlighted. Using a 10× objective lens ($r_0 \approx 7$ μm), the following frequency-dependent strategies were adopted: a low modulation frequency (~100-200 Hz) is used to determine $k_r/(Cr_0^2)$ and $\sqrt{k_zC}/(hC)$, a moderate frequency (~100 kHz) isolates $k_{r,Al}/(C_{Al}r_0^2)$, and higher frequencies (~0.4-1 MHz) are employed to determine $\sqrt{k_zC}/(h_{Al}C_{Al})$.

Figures 2(c-h) present the measured SPS signals of the Kapton HN-100 PI film at modulation frequencies of 0.6 MHz, 90 kHz, and 150 Hz, along with their corresponding sensitivity analyses. For clarity in visualization, only selected portions of the time-domain data are displayed. However,

the full periodic heating cycles were used in the fitting process for parameter extraction, as illustrated in the insets of Figures. 2(c), 2(e), and 2(g).

Figure 2(d) shows that at 0.6 MHz, the signals are sensitive to both $k_{r,\text{Al}}/(C_{\text{Al}}r_0^2)$ and $\sqrt{k_z C}/(h_{\text{Al}} C_{\text{Al}})$, with strong parameter correlation. At 90 kHz, sensitivity is dominated by $k_{r,\text{Al}}/(C_{\text{Al}}r_0^2)$, as shown in Fig. 2(f). Therefore, joint fitting of these datasets allows accurate determination of both composite parameters. With known values of $h_{\text{Al}}$, $C_{\text{Al}}$, and $r_0$, the individual values of $k_{r,\text{Al}}$ and $\sqrt{k_z C}$ can be extracted.

At 150 Hz (Fig. 2(g, h)), the signals are sensitive to both $k_r/(Cr_0^2)$ and $\sqrt{k_z C}/(hC)$, with uncoupled sensitivities. Figure 2(h) shows that the magnitude of the sensitivity coefficient to $h$ ($S_h$) increases steadily from 0 to 0.6 over the normalized time range of 0.5-0.8, while the sensitivity coefficient to $k_r$ ($S_{k_r}$) remains nearly constant at around 0.4. This suggests that $\sqrt{k_z C}/(hC)$ mainly affects the slope of the signals, while $k_r/(Cr_0^2)$ primarily affects the signal amplitude within this time window. Consequently, both quantities can be extracted simultaneously by fitting the low-frequency dataset. With $r_0$ and $\sqrt{k_z C}$ already determined, values of $k_r/C$ and $hC$ are obtained. Since $h$ is known, the final individual properties of $k_{r,\text{Al}}$, $k_r$, $k_z$, and $C$ are determined.

For Kapton HN-100 PI film, the extracted thermal properties are: $k_{r,\text{Al}} = 47 \pm 2$ W m$^{-1}$ K$^{-1}$, $k_r = 0.56 \pm 0.04$ W m$^{-1}$ K$^{-1}$, $k_z = 0.175 \pm 0.02$ W m$^{-1}$ K$^{-1}$, and $C = 1.74 \pm 0.09$ MJ m$^{-3}$ K$^{-1}$. The uncertainties of the fitted parameters were estimated by propagating errors through the full covariance matrix, derived using the Jacobian of the model with respect to the parameters. For further details of the methodology, we refer the reader to Refs. [21,27,28].

To independently verify $k_{r,\text{Al}}$, we measured the electrical resistance of the Al film using the van der Pauw method and converted it to thermal conductivity via the Wiedemann-Franz law, assuming a Lorentz number of $L = 2.14 \times 10^{-8}$ W Ω K$^{-2}$ [29]. This yields the electronic contribution to the in-plane thermal conductivity of the Al film as $k_{r,\text{Al}}^{(e)} = 42$ W m$^{-1}$ K$^{-1}$, slightly lower than the total $k_{r,\text{Al}}$ obtained from our SPS measurements. The difference is attributed to the phonon contribution to the Al film's thermal conductivity. This agreement supports the reliability of our SPS-based measurements.

Similar measurements were conducted for the other suspended PI films, and the results are presented in Table 1.

*3.2. Thermal characterization of spin-coated PI films*

Three PI films with different thicknesses were spin-coated onto fused silica substrates. After thermal curing, their thicknesses were measured using a step profiler, yielding values of 4.45 ± 0.1 μm, 5.0 ± 0.1 μm, and 10 ± 0.1 μm. Subsequently, an Al film was deposited onto each PI film as a transducer layer. The sample structure is illustrated in Fig. 4(a), encompassing 12 parameters: $k_{Al}$, $C_{Al}$, and $h_{Al}$ of the Al layer; $k_r$, $k_z$, $C$, and $h$ of the PI film; the interfacial thermal conductance $G_1$ between Al and PI and $G_2$ between PI and silica; the thermal conductivity and volumetric specific heat capacity of fused silica ($k_{silica}$ and $C_{silica}$); and the laser spot radius ($r_0$).

Among these parameters, $G_1$ and $G_2$ have negligible influence on the measurement signals and are therefore treated as fixed inputs, with values set to $G_1 = G_2 = 100 \pm 50$ MW m$^{-2}$ K$^{-1}$. The following six parameters – $C_{Al}$, $h_{Al}$, $h$, $k_{silica}$, $C_{silica}$, and $r_0$ – must be accurately predetermined to serve as reliable inputs. $C_{Al}$, $h_{Al}$, $h$, and $r_0$ are obtained using the method described previously, while $k_{silica}$ and $C_{silica}$ are adopted from literature values [25,26], as discussed in Section 3.1.

The remaining four parameters – $k_{Al}$, $k_r$, $k_z$, and $C$ – are simultaneously determined through multi-frequency SPS measurements. Same as in the case of suspended film structure, these parameters are embedded within four composite quantities: $k_{r,Al}/(C_{Al}r_0^2)$, $k_r/(Cr_0^2)$, $\sqrt{k_zC}/(h_{Al}C_{Al})$, and $\sqrt{k_zC}/(hC)$. By fitting these composite parameters and combining them with the predetermined input values, the individual thermal properties of the Al layer and PI films can be simultaneously extracted.

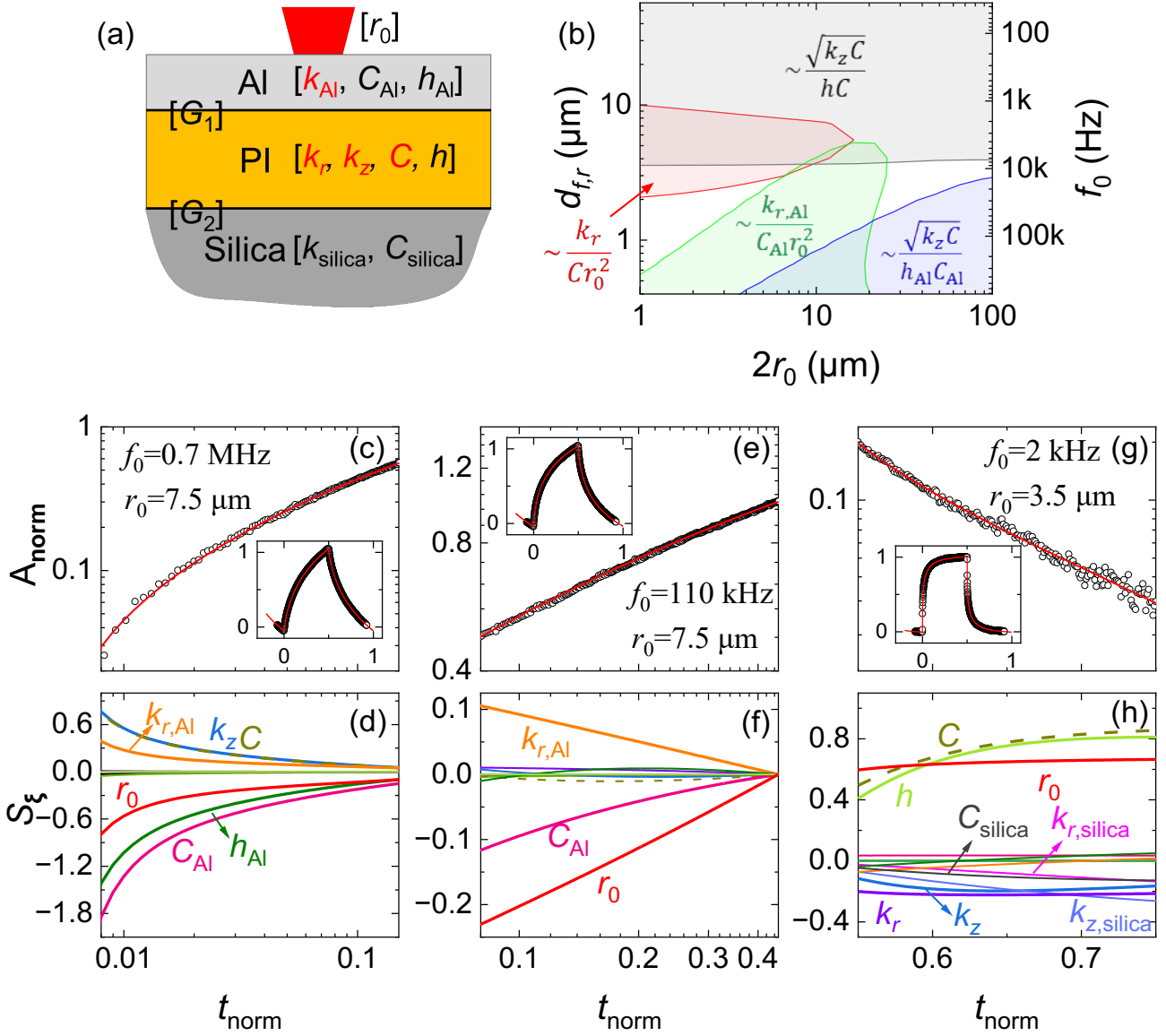

**Fig. 4.** (a) Sample structure and thermal model for measuring thermal properties of spin-coated PI films. (b) Contour plot of sensitivity coefficients for $k_{r,\mathrm{Al}}/(C_{\mathrm{Al}}r_0^2)$, $\sqrt{k_zC}/(h_{\mathrm{Al}}C_{\mathrm{Al}})$, $k_r/(Cr_0^2)$ and $\sqrt{k_zC}/(hC)$ as functions of laser spot diameter and in-plane thermal diffusion length. Regions with sensitivity magnitudes greater than 0.2 are highlighted. (c-h) Measured SPS signals of the 5-μm thick PI film at modulation frequencies of (c, d) 0.7 MHz, (e, f) 0.11 MHz, and (g,h) 2 kHz, along with corresponding sensitivity analyses for key parameters. The insets in panels (c), (e), and (g) show the fitting of the signals over the full periodic heating cycle.

Due to the influence of the fused silica substrate, heat diffusion within the supported PI films – and thus the sensitivity of the SPS signals – differs from that of suspended films. To optimize the experimental conditions for accurate determination of the composite parameters, Fig. 4(b) presents the contour of sensitivity coefficients of the four composite parameters for the case of supported PI films as functions of $2r_0$ and $d_{f,r}$.

As shown in Fig. 4(b), the conductive substrate enhances cross-plane heat diffusion, resulting in high sensitivity to $\sqrt{k_z C}/(hC)$ at modulation frequencies below 10 kHz. To accurately determine $k_r/(Cr_0^2)$, a small laser spot size ($2r_0 \approx 7$ μm, corresponding to a 20× objective lens) and an intermediate modulation frequency (~1-2 kHz) are preferred. At high modulation frequencies (>100 kHz), $k_{r,\text{Al}}/(C_{\text{Al}} r_0^2)$ can be determined using a small spot size (20× or 10× objective lens), while $\sqrt{k_z C}/(h_{\text{Al}} C_{\text{Al}})$ can be extracted using a larger spot size (10× or 5× objective lens).

Detailed measurements of the 5 μm spin-coated PI film are presented below. Figures 4(c)-4(h) show the measured SPS signals of the 5 μm spin-coated PI film at modulation frequencies of 0.7 MHz, 110 kHz, and 2 kHz, along with corresponding sensitivity analyses. The first two datasets were acquired using a spot size of $r_0 = 7.5$ μm, while the last one was measured with a reduced spot size of $r_0 = 3.5$ μm.

Similar to the case of suspended PI film, Fig. 4(d) shows that the 0.7 MHz signals are sensitive to both $k_{r,\text{Al}}/(C_{\text{Al}} r_0^2)$ and $\sqrt{k_z C}/(h_{\text{Al}} C_{\text{Al}})$, with significant coupling between these two parameters. At a lower modulation frequency of 110 kHz, the signals become predominantly sensitive to $k_{r,\text{Al}}/(C_{\text{Al}} r_0^2)$, as shown in Fig. 4(f). Therefore, these two datasets allow the determination of $k_{r,\text{Al}}$ and $\sqrt{k_z C}$. Furthermore, measurements at 2 kHz with a reduced spot size enable the determination of $k_r/(Cr_0^2)$ and $\sqrt{k_z C}/(hC)$, following the same rationale as for the suspended PI film.

Simultaneously fitting these three datasets yields the thermal properties of the Al and supported PI film as: $k_{r,\text{Al}} = 42 \pm 2$ W m$^{-1}$ K$^{-1}$, $k_r = 0.55 \pm 0.12$ W m$^{-1}$ K$^{-1}$, $k_z = 0.32 \pm 0.02$ W m$^{-1}$ K$^{-1}$, and $C = 1.53 \pm 0.09$ MJ m$^{-3}$ K$^{-1}$. Similar measurements were also performed on the other spin-coated PI films, and the results are presented in Table 1.

*3.3. Discussion and comparison with literature*

The measured thermal properties of the six PI films are summarized in Table 1, with comparisons to literature values illustrated in Fig. 5. Specifically, Fig. 5(a) shows $k_r$ and $k_z$, while Fig. 5(b) presents the measured $C$. For comparison with literature data, the thermal diffusivities reported by Takahashi et al.[13] were converted to thermal conductivities by assuming a constant $C$ of 1.61 MJ m$^{-3}$ K$^{-1}$.

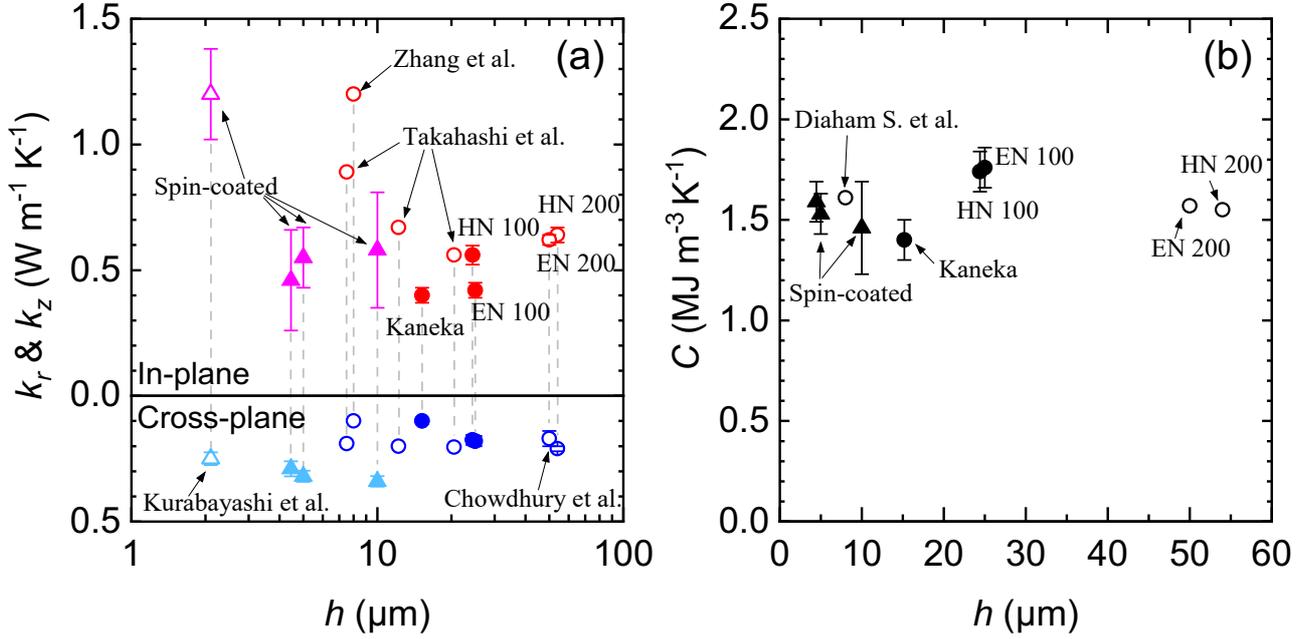

**Fig. 5.** (a) Measured thermal conductivities compared with literature values; (b) Measured volumetric heat capacities compared with literature values. Solid symbols represent results from this work, while open symbols correspond to literature data. Triangles indicate spin-coated PI films, and circles represent suspended commercial PI films.

Our measurements indicate that both suspended and spin-coated PI films exhibit similar $k_r$, ranging from 0.4 to 0.6 W m$^{-1}$ K$^{-1}$, although the spin-coated PI films show larger uncertainties due to the reduced measurement sensitivity. Notably, the spin-coated PI films exhibit $k_z$ approximately twice that of the suspended PI films, resulting in a lower anisotropy ratio ($\eta$=1.6~1.7) compared to the suspended PI films ($\eta \approx$ 2.3-4).

For commercial suspended PI films, the measured values for $k_r$, $k_z$, and $\eta$ for Kapton HN-100 and EN-100 (25 μm thick) agree well with those reported by Chowdhury et al.[15] for Kapton HN-200 and EN-200 films (50 μm thick). These results are also consistent with those of a 20-μm-thick UPILEX-R film reported by Takahashi et al. [10]. However, the increase in $k_r$ with decreasing $h$ observed by Takahashi et al. [10] was not observed in our measurements, likely due to differences in material sources, as our 15 μm and 25 μm films were obtained from different manufacturers. Takahashi et al. [10] also reported an increase in anisotropy ratio from 2.7 (20.5 μm thick) to 5 (7.5 μm thick) for UPILEX-R films. In contrast, both our measurements and those of Chowdhury et al.[15] show anisotropy ratios ranging from 2.3 to 4. However, Zhang et al. [14] reported a notably high anisotropy ratio of 12 for an 8-μm-thick PI film, but did not specify the polyimide type, the presence or absence of a substrate, or any processing parameters such as curing temperature or spin-coating conditions. Additionally, their thermal analysis relied on a literature value for $C$ without experimental validation. These omissions reduce the ability to interpret or reproduce their results and emphasize the importance of reporting full material details and methodological assumptions when characterizing thermal transport in polymer films.

For the spin-coated PI films, our measured $k_z$ values are in good agreement with those reported by Kurabayashi et al. [9]. However, Kurabayashi et al. [9] reported $k_r$ values for a 2-μm-thick spin-coated PI film that were more than twice those obtained in our study. This discrepancy is likely attributable to differences in the polyimide precursor solution, spin-coating parameters, thermal curing protocols, or even differences in the thermal measurement techniques employed.

The observed enhancement in $k_z$ of spin-coated PI films, compared to their suspended commercial counterparts, can be attributed to several morphological and interfacial factors. First, film morphology—particularly the density and continuity of polymer chains along the cross-plane direction—plays a significant role. Spin-coated films often undergo thermal curing, which can lead to denser packing and improved chain stacking perpendicular to the substrate, thereby facilitating phonon transport across layers. Second, inter-chain alignment induced during spin coating and annealing may result in partial vertical ordering, especially when strong shear or solvent evaporation gradients are present during processing. This contrasts with commercial Kapton films, which are typically biaxially stretched and exhibit pronounced in-plane chain alignment, thereby reducing phonon pathways in the cross-plane direction. Finally, substrate influence cannot be overlooked. The rigid fused silica substrate used for spin-coated films may constrain polymer relaxation during curing, suppressing void formation and improving interfacial adhesion. This intimate film–substrate contact can reduce thermal boundary resistance and enhance heat conduction from the metal transducer into the underlying PI film. Collectively, these effects may account for the elevated cross-plane thermal conductivities and reduced anisotropy ratios observed in spin-coated samples.

To verify the difference in anisotropy between the commercial and spin-coated films as described above, we performed polarized Raman spectroscopy experiments on both samples. Both films exhibit a Raman peak near 1390 cm$^{-1}$. The peak at this position was fitted, and the results are shown in Fig. 6. To enable a quantitative comparison, we introduced the depolarization ratio $\rho$, which is defined as follows [30]:

$$\rho = \frac{I_\perp}{I_\parallel} \tag{5}$$

where $I_\perp$ represents the intensity of Raman scattered light polarized perpendicular to the polarization plane of the incident light, while $I_\parallel$ represents the intensity of Raman scattered light polarized parallel to the polarization plane of the incident light. The calculated depolarization ratio for the commercial film is $\rho_1 = 0.60$, while that for the spin-coated film is $\rho_2 = 0.71$, demonstrating that the spin-coated film indeed exhibits a more isotropic nature.

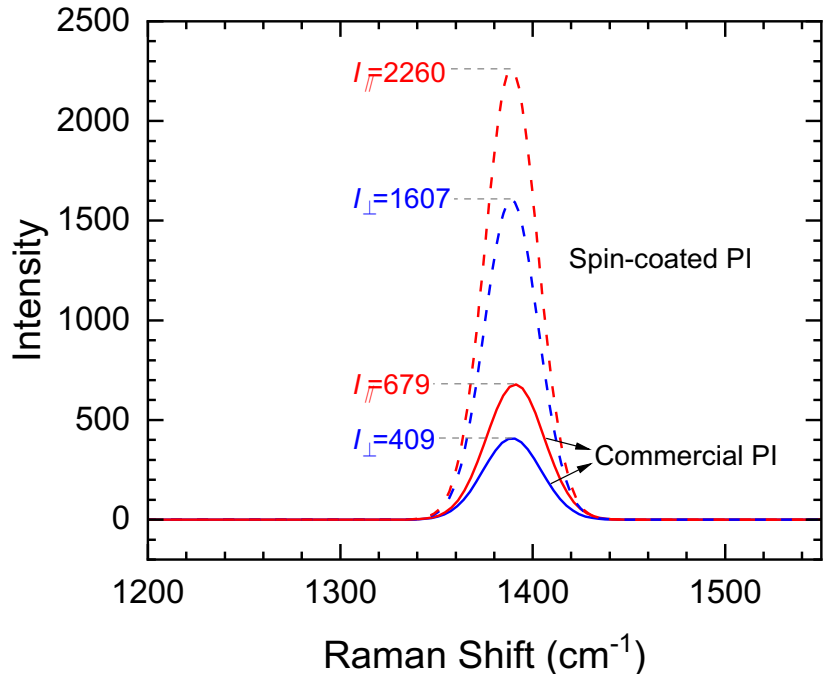

**Fig. 6.** Raman polarization spectra of commercial and spin-coated PI films. The solid lines represent the commercial film, while the dashed lines represent the spin-coated film. Red lines correspond to the polarization of the scattered light parallel to the incident light polarization plane, and blue lines correspond to the polarization perpendicular to the incident light polarization plane.

Our measured $C$ values for both suspended and spin-coated PI films range from 1.4 to 1.76 MJ m$^{-3}$ K$^{-1}$, are consistent with the widely accepted value of 1.613 MJ m$^{-3}$ K$^{-1}$ reported by Diaham et al. [17]. A closer scrutiny reveals that both Kapton films have a higher heat capacity of 1.76 MJ m$^{-3}$ K$^{-1}$, whereas the Kaneka film has a lower heat capacity value of 1.4 MJ m$^{-3}$ K$^{-1}$, representing a discrepancy of up to 25% between the two brands. In contrast, the heat capacity values of the spin-coated PI films show closer agreement with the literature value.

**Table 1:** Measurement results for PI films.

| PI films | Thickness μm | In-plane $k_r$ W m$^{-1}$ K$^{-1}$ | Cross-plane $k_z$ W m$^{-1}$ K$^{-1}$ | Anisotropic ratio $\eta = k_r/k_z$ | Heat capacity $C$ MJ m$^{-3}$ K$^{-1}$ |
|---|---|---|---|---|---|
| Kaneka | 15.2 ± 0.2 | 0.4 ± 0.03 | 0.1 ± 0.01 | 4 | 1.4 ± 0.10 |
| Kapton HN-100 | 24.4 ± 0.2 | 0.56 ± 0.04 | 0.175 ± 0.02 | 3.2 | 1.74 ± 0.09 |
| Kapton EN-100 | 25.0 ± 0.2 | 0.42 ± 0.03 | 0.18 ± 0.02 | 2.33 | 1.76 ± 0.11 |
| Spin-coated | 4.45 ± 0.1 | 0.46 ± 0.2 | 0.29 ± 0.03 | 1.58 | 1.59 ± 0.09 |
| Spin-coated | 5.0 ± 0.1 | 0.55 ± 0.12 | 0.32 ± 0.02 | 1.71 | 1.53 ± 0.09 |
| Spin-coated | 10.0 ± 0.1 | 0.58 ± 0.23 | 0.34 ± 0.02 | 1.71 | 1.46 ± 0.24 |

In summary, although the measured thermal conductivity and volumetric heat capacity vary among different types of PI films, the thermal anisotropy ratio generally lies within the range of 1.58 to 4. For spin-coated PI films, the measurement results are highly sensitive to the quality of the polyimide precursor solution and the thermal curing conditions. In this study, the spin-coated films demonstrated a distinct enhancement in $k_z$ accompanied by a reduced thermal anisotropy ratio compared to the commercial PI films.

## 4. CONCLUSIONS

In this work, we employed the SPS method to characterize three commercial PI films and three spin-coated PI films on fused silica substrates. The obtained thermal conductivity, specific heat capacity, and thermal anisotropic ratio show good agreement with literature values, thereby establishing a novel approach for measuring similar polymer thin films. Notably, our methodology eliminates the need to use volumetric heat capacity as a predefined input parameter, a key advantage that represents a significant advancement over previous studies. Instead, these values are directly determined from the measurements themselves, thereby avoiding potential error propagation inherent in conventional approaches. Looking ahead, the versatility of the SPS technique positions it as a valuable tool for investigating thermal transport in complex polymer systems, including composites, nano-layered structures, and materials designed for flexible electronic applications.

**ACKNOWLEDGMENT**

This work is supported by the National Natural Science Foundation of China (NSFC) through Grant No. 52376058.

**DATA AVAILABILITY**

The data that support the findings of this study are available from the corresponding author upon reasonable request.

**REFERENCES**

[1]  H. Ohya, Polyimide membranes: Applications, fabrications and properties, CRC Press LLC, Milton, 1997.
[2]  K. Tao, G. Sun, C. Feng, G. Liu, Y. Li, R. Chen, J. Wang, S. Han, Structural Design and Research Progress of Thermally Conductive Polyimide Film – A Review, Macromol. Rapid Commun. 44 (2023) 2300060. https://doi.org/10.1002/marc.202300060.
[3]  H. Chen, V.V. Ginzburg, J. Yang, Y. Yang, W. Liu, Y. Huang, L. Du, B. Chen, Thermal conductivity of polymer-based composites: Fundamentals and applications, Progress in Polymer Science 59 (2016) 41–85. https://doi.org/10.1016/j.progpolymsci.2016.03.001.
[4]  A.B. Frazier, Recent applications of polyimide to micromachining technology, IEEE Trans. Ind. Electron. 42 (1995) 442–448. https://doi.org/10.1109/41.464605.
[5]  A. Sezer Hicyilmaz, A. Celik Bedeloglu, Applications of polyimide coatings: A review, SN Appl. Sci. 3 (2021) 363. https://doi.org/10.1007/s42452-021-04362-5.
[6]  C.L. Choy, Thermal conductivity of polymers, Polymer 18 (1977) 984–1004. https://doi.org/10.1016/0032-3861(77)90002-7.
[7]  R.J. Samuels, N.E. Mathis, Orientation Specific Thermal Properties of Polyimide Film, J. Electron. Packag. 123 (2001) 273–277. https://doi.org/10.1115/1.1347986.
[8]  D. Yorifuji, S. Ando, Molecular Structure Dependence of Out-of-Plane Thermal Diffusivities in Polyimide Films: A Key Parameter for Estimating Thermal Conductivity of Polymers, Macromolecules 43 (2010) 7583–7593. https://doi.org/10.1021/ma101066p.
[9]  K. Kurabayashi, M. Asheghi, M. Touzelbaev, K.E. Goodson, Measurement of the thermal conductivity anisotropy in polyimide films, J. Microelectromech. Syst. 8 (1999) 180–191. https://doi.org/10.1109/84.767114.
[10] Q. Zheng, D. Chalise, M. Jia, Y. Zeng, M. Zeng, M. Saeidi-Javash, A.N.M. Tanvir, G. Uahengo, S. Kaur, J.E. Garay, T. Luo, Y. Zhang, R.S. Prasher, C. Dames, Structured illumination with thermal imaging (SI-TI): A dynamically reconfigurable metrology for parallelized thermal transport characterization, Applied Physics Reviews 9 (2022) 021411. https://doi.org/10.1063/5.0079842.

[11] D. Wang, H. Ban, P. Jiang, Spatially resolved lock-in micro-thermography (SR-LIT): A tensor analysis-enhanced method for anisotropic thermal characterization, Applied Physics Reviews 11 (2024) 021407. https://doi.org/10.1063/5.0191073.

[12] D. Wang, H. Ban, P. Jiang, Three-dimensional (3D) tensor-based methodology for characterizing 3D anisotropic thermal conductivity tensor, International Journal of Heat and Mass Transfer 242 (2025) 126886. https://doi.org/10.1016/j.ijheatmasstransfer.2025.126886.

[13] F. Takahashi, K. Ito, J. Morikawa, T. Hashimoto, I. Hatta, Characterization of Heat Conduction in a Polymer Film, Jpn. J. Appl. Phys. 43 (2004) 7200–7204. https://doi.org/10.1143/JJAP.43.7200.

[14] Y. Zhang, R. Xu, Y. Liu, Q. Jiang, Q. Li, Y. Liu, J. Wang, Anisotropic thermal diffusivity measurement of thin films: From a few to hundreds of microns, International Journal of Heat and Mass Transfer 227 (2024) 125536. https://doi.org/10.1016/j.ijheatmasstransfer.2024.125536.

[15] N. Chowdhury, J. Sun, D.G. Cahill, Anisotropic thermal conductivity of kapton films, composites, and laminates, ACS Appl. Polym. Mater. (2025) acsapm.4c03181. https://doi.org/10.1021/acsapm.4c03181.

[16] B. Kim, Y. Lee, D. Yang, K. Fushinobu, Y. Kim, M. Nomura, Rapid electro-thermal micro-actuation of flat optics enabled by laser-induced graphene on colorless polyimide substrates, Advanced Optical Materials 13 (2025) 2500498. https://doi.org/10.1002/adom.202500498.

[17] S. Diaham, F. Saysouk, M. Locatelli, B. Belkerk, Y. Scudeller, R. Chiriac, F. Toche, V. Salles, Thermal conductivity of polyimide/boron nitride nanocomposite films, J of Applied Polymer Sci 132 (2015) app.42461. https://doi.org/10.1002/app.42461.

[18] P. Jiang, X. Qian, R. Yang, Time-domain thermoreflectance (TDTR) measurements of anisotropic thermal conductivity using a variable spot size approach, Review of Scientific Instruments 88 (2017) 074901. https://doi.org/10.1063/1.4991715.

[19] A.J. Schmidt, R. Cheaito, M. Chiesa, A frequency-domain thermoreflectance method for the characterization of thermal properties, Review of Scientific Instruments 80 (2009) 094901. https://doi.org/10.1063/1.3212673.

[20] S. Song, T. Chen, P. Jiang, Comprehensive thermal property measurement of semiconductor heterostructures using the square-pulsed source (SPS) method, Journal of Applied Physics 137 (2025) 055101. https://doi.org/10.1063/5.0244681.

[21] T. Chen, S. Song, Y. Shen, K. Zhang, P. Jiang, Simultaneous measurement of thermal conductivity and heat capacity across diverse materials using the square-pulsed source (SPS) technique, International Communications in Heat and Mass Transfer 158 (2024) 107849. https://doi.org/10.1016/j.icheatmasstransfer.2024.107849.

[22] P. Jiang, D. Wang, Z. Xiang, R. Yang, H. Ban, A new spatial-domain thermoreflectance method to measure a broad range of anisotropic in-plane thermal conductivity, International Journal of Heat and Mass Transfer 191 (2022) 122849. https://doi.org/10.1016/j.ijheatmasstransfer.2022.122849.

[23] T. Chen, P. Jiang, Decoupling thermal properties in multilayered systems for advanced thermoreflectance experiments, Phys. Rev. Applied 23 (2025) 044004. https://doi.org/10.1103/PhysRevApplied.23.044004.

[24] Y.S. Touloukian, E.H. Buyco, Thermophysical properties of matter - the TPRC data series. Volume 4. Specific heat - metallic elements and alloys. (reannouncement). Data book, Purdue Univ., Lafayette, IN (United States). Thermophysical and Electronic Properties Information Center, 1971. https://www.osti.gov/biblio/5439707 (accessed February 5, 2025).

[25] A. Sugawara, The precise determination of thermal conductivity of pure fused quartz, Journal of Applied Physics 39 (1968) 5994–5997. https://doi.org/10.1063/1.1656103.

[26] R. Brückner, Properties and structure of vitreous silica. I, Journal of Non-Crystalline Solids 5 (1970) 123–175. https://doi.org/10.1016/0022-3093(70)90190-0.

[27] M. Zhang, T. Chen, S. Song, Y. Bao, R. Guo, W. Zheng, P. Jiang, R. Yang, Extending the low-frequency limit of time-domain thermoreflectance via periodic waveform analysis, Journal of Applied Physics 138 (2025) 055101. https://doi.org/10.1063/5.0275018.

[28] J. Yang, E. Ziade, A.J. Schmidt, Uncertainty analysis of thermoreflectance measurements, Review of Scientific Instruments 87 (2016) 014901. https://doi.org/10.1063/1.4939671.

[29] AshcroftSolidState.pdf, (n.d.). https://ia800503.us.archive.org/25/items/AshcroftSolidState/AshcroftSolidState.pdf (accessed May 17, 2025).

[30] D.K. Singh, A. Kumar Mishra, A. Materny, eds., Raman spectroscopy: Advances and applications, Springer Nature Singapore, Singapore, 2024. https://doi.org/10.1007/978-981-97-1703-3.